\documentclass[aps prapplied, twocolumn, superscriptaddress, longbibliography, floatfix]{revtex4-1}

\usepackage{graphicx}
\usepackage{amsmath}
\usepackage{bm}
\usepackage{color,soul}
\usepackage{gensymb}
\usepackage[pagewise]{lineno}

\definecolor{mygreen}{rgb}{0.0, 0.7, 0.2}

\begin{document}
\setlength{\textfloatsep}{12pt}

\title{Bridging high-\textit{Q} and Kerr-nonlinear photonics using modal phase matching}


\author{Jordan R. Stone}
\email{jordan.stone@nist.gov. This document is preliminary and is intended for peer review conducted by a journal or conference.}
\affiliation{National Institute for Standards and Technology, Gaithersburg, MD 20899}

\author{Saleha Fatema}
\affiliation{National Institute for Standards and Technology, Gaithersburg, MD 20899}
\affiliation{Whiting School of Engineering, Johns Hopkins University, Baltimore, MD 21218}

\author{Christopher V. Poulton}
\affiliation{Beacon Photonics, 3000 Wilson Blvd., Arlington, VA 22201}

\author{Michael G. Wood}
\affiliation{Beacon Photonics, 3000 Wilson Blvd., Arlington, VA 22201}

\author{Gordon A. Keeler}
\affiliation{Beacon Photonics, 3000 Wilson Blvd., Arlington, VA 22201}

\author{Kartik Srinivasan}
\affiliation{National Institute for Standards and Technology, Gaithersburg, MD 20899}
\affiliation{Joint Quantum Institute, NIST/University of Maryland, College Park, MD 20742}


\date{\today}

\begin{abstract}
Integrated Kerr microresonators provide on-chip optical nonlinearity for wavelength conversion, optical frequency combs, and quantum light sources. Traditionally, their dispersion engineering has tied nonlinear functionality to resonator geometry, forcing trade-offs with other device objectives. In particular, Kerr microresonators usually feature narrow resonator waveguides, but wide waveguides support higher $Q$ through reduced sidewall scattering. We propose that modal phase matching - invoking multiple spatial mode families to satisfy dispersion requirements - facilitates Kerr nonlinear optics beyond traditional geometries. Working with a commercial foundry, we design and fabricate high-$Q$ ($>10^7$) microresonators on a 160-nm-thick silicon nitride platform and demonstrate Kerr optical parametric oscillation. We achieve $20$\% conversion efficiency and gap-free wavelength tuning over \(>1\) nm for parametric oscillation at the cesium D$_1$ transition. Modal phase matching further supports pumping in both $1060$-nm and $795$-nm bands, without custom device layers, for wavelength generation between 600~nm to 1400~nm. Our work expands the Kerr design space, effectively decoupling $Q$ and dispersion to create new opportunities with high-$Q$ nonlinear devices.  
\end{abstract}


\maketitle



Integrated photonics research constantly pushes the limits of nanofabrication and device physics, advancing our understanding of nanophotonic systems and facilitating important prototype demonstrations \cite{pelucchi2022potential, price2022roadmap, miri2019exceptional, Du2023moire, zheng2026large, sipahigil2016integrated, spencer2018optical, yang2025nanofabrication}. It is also true that prototypes are most disruptive when they are compatible with wafer-scale integration, at the foundry level, that scales production and drives down costs. For example, in photonic integrated circuits based on planar silicon nitride (Si$_3$N$_4$, hereafter SiN) waveguides, this often entails thin waveguide layers ($<400$ nm) buried in SiO$_2$ cladding, and adhering to minimum feature/gap sizes \cite{munoz2019foundry, zhang2024300}. The SiN platform is ideal for wafer-scale integration, appealing for its low optical losses and large Kerr nonlinearity \cite{xiang2022silicon, moss_new_2013}.  

Light sources based on Kerr nonlinearity, including optical frequency combs, optical parametric oscillators, third harmonic generators, photon pair sources, and quantum frequency converters, exhibit rich dynamics and high-end performance, with research spanning fundamental nonlinear physics to prototype development and application demonstrations \cite{moille2023kerr, jang2019observation, kim2021synchronization, newman2019architecture, reimer2016generation, surya2018efficient, long2024sub, raghunathan2025telecom}. In this context, there is an emerging emphasis on foundry fabrication of Kerr microresonators, with recent demonstrations of frequency combs and optical parametric oscillation (OPO) in foundry SiN platforms \cite{ou2025300, zang2024foundry, liu2025implementing, ye2023foundry, yuan2023soliton, liu2025near}. Such Kerr nonlinear photonics typically use tall, narrow waveguides, often air-clad on the sides and top, to control chromatic dispersion. This presents three major problems. First, tall waveguides require elaborate deposition, etching, and annealing techniques to mitigate stress \cite{luke2013overcoming, bose2024anneal}, making co-integration with other photonics/electronics layers challenging. Second, such designs run counter to geometries that minimize scattering loss in pursuit of resonators with high intrinsic quality factors ($Q$), which can enhance nonlinear efficiency or serve as stable frequency references \cite{liu202236}. Lastly, conventional dispersion engineering constrains the pump wavelength; hence, different nonlinear processes or wavelength targets require multiple device layer thicknesses. Although recent work has demonstrated narrowband frequency combs in coupled SiN microresonators on a high-$Q$ foundry platform \cite{yuan2023soliton, liu2025near}, broadband dispersion engineering is required to nonlinearly couple widely-separated wavelengths for OPO, quantum frequency conversion, etc. 
\begin{figure*}[t!]
\centering
\includegraphics[width=\linewidth]{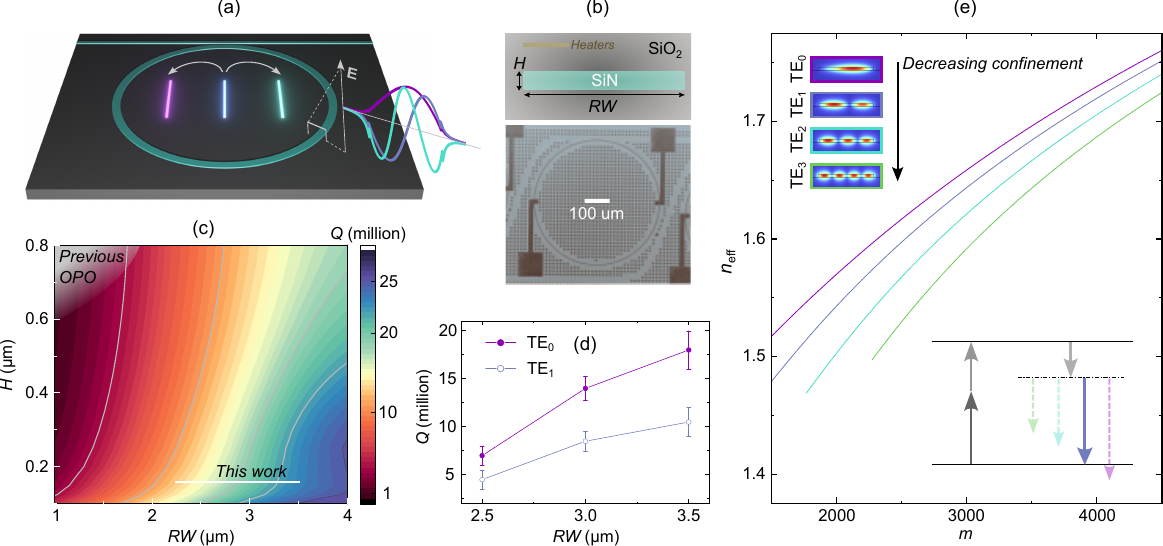}
\caption{\textbf{Modal phase matching unites high-$Q$ and Kerr-nonlinear geometries in a foundry silicon nitride (SiN) platform.} (a) In this platform, a SiN microring resonator can support multiple high-$Q$ mode families and their Kerr nonlinear interactions. (b) Top: Schematic of the resonator waveguide cross section, with height $H$ and width $RW$; bottom: Resonator imaged from above. (c) Simulated $Q$ versus $H$ and $RW$, with ring radius $= 250$ $\mu$m, for the fundamental transverse-electric (TE$_0$) mode near $1060$ nm. Solid gray lines follow contours of constant $Q$. (d) Measurements of $Q$ versus $RW$. Error bars cover the range of values measured in the $1060$ nm band on multiple devices. (e) Simulated effective index, $n_{\rm{eff}}$, versus azimuthal mode number, $m$, of TE$_0$, TE$_1$, TE$_2$, and TE$_3$ modes in a microresonator with $RW = 3$ $\mu$m and $H = 160$ nm. The range of $m$ values corresponds approximately to wavelengths between $600$ nm and $1590$ nm. Differences in $n_{\rm{eff}}$ are caused by differences in modal confinement, providing a design principle for four-wave mixing (FWM), illustrated by the inset.}\label{fig:one}
\end{figure*}

Here, we report on Kerr OPO in 160-nm-thick, SiO$_2$-clad, high-$Q$ SiN microresonators fabricated by a commercial foundry. Modal phase matching, so-called hybrid-mode OPO \cite{zhou2022hybrid}, is used to satisfy dispersion requirements, and the OPO signal is tunable across visible and near-infrared wavelengths according to the mode configuration, pump wavelength, and resonator dimensions. The intrinsic $Q$ of the fundamental transverse-electric (TE$_0$) mode approaches $20$ million at 1060 nm wavelength and $30$ million at 1300 nm. Moreover, simulations indicate that improved surface roughness (at levels reported by commercial foundries) would allow $Q>50$ million ($100$ million) at $1060$ nm ($1300$ nm). We explore the design considerations, conversion efficiency, and tuning characteristics of a hybrid-mode OPO pumped near $1060$ nm and targeting the Cs D$_1$ transition near $894.6$ nm. On-chip heaters enable gap-free tuning of the signal wavelength between 894.5 nm and 895.9 nm without sacrificing efficiency. We also pump near $795$ nm to generate signal tones near $700$ nm or $640$ nm; traditionally, different pump bands require custom device layers. Modal phase matching, which is established in $\chi^{(2)}$ platforms \cite{zhu2021integrated, chen2018modal, luo2019optical}, can be applied to most other Kerr nonlinear photonics; we anticipate that high-$Q$ foundry platforms for OPO, quantum frequency conversion, third harmonic generation, and photon pair sources are immediately within reach.   

\section{\label{sec:design}Design}

\begin{figure*}[t!]
\centering
\includegraphics[width=\linewidth]{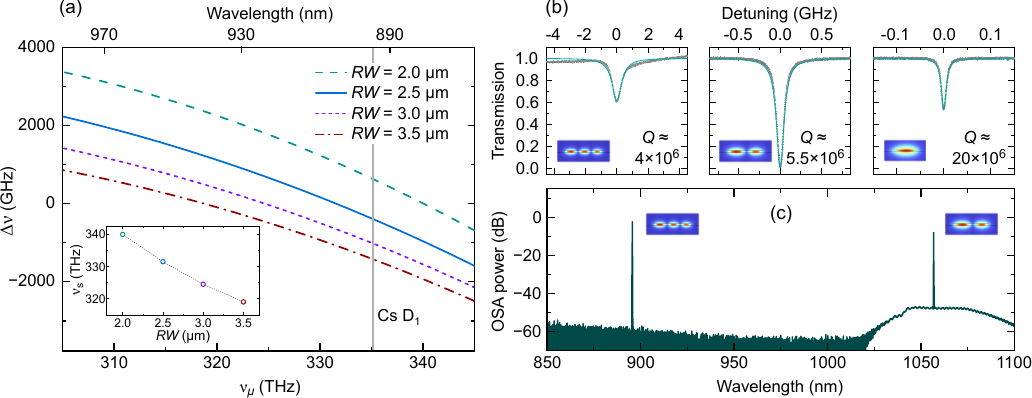}
\caption{\textbf{Design and characterization of a hybrid-mode Kerr optical parametric oscillator (OPO) for the Cs D$_1$ transition.} (a) Simulated frequency mismatch, $\Delta\nu$, spectrum of a TE$_0$-TE$_1$-TE$_2$ OPO for various $RW$ values. The inset shows the OPO signal frequency, $\nu_{\rm{s}} = \nu_{\rm{\mu}}[\Delta\nu = 0]$ versus $RW$. (b) Normalized transmitted power versus frequency for a TE$_2$ mode near 895 nm (left), a TE$_1$ mode near 1060 nm (center), and a TE$_0$ mode near 1300 nm, corresponding to the OPO signal, pump, and idler modes, respectively. (c) OPO spectrum, zoomed in to show only the pump and signal tones, recorded with an optical spectrum analyzer (OSA). $0$ dB is referenced to $1$ mW. We measure $\approx18\times$ loss between the output waveguide and OSA, used to convert the OSA power to on-chip power.}\label{fig:two}
\end{figure*}

Figure \ref{fig:one} illustrates the key relationships between microresonator geometry (a-b), $Q$ (c-d), and dispersion (e) that motivate our study. A prominent source of loss is scattering induced by surface roughness. Hence, $Q$ depends on the overlap between optical modes and the resonator surface \cite{corato2024absorption}. In Fig. \ref{fig:one}(c), we simulate $Q$ at 1060 nm for the TE$_0$ mode in SiO$_2$-clad SiN microresonators with $250$ $\mu$m ring radius, versus the microresonator waveguide width, $RW$, and height, $H$, that are denoted in Fig. \ref{fig:one}(b) (see Supplemental Material for details about the modeling). In general, $Q$ increases with $RW$ due to stronger modal confinement; however, its dependence on $H$ is more nuanced. On the one hand, increasing $H$ also leads to stronger modal confinement; on the other hand, it increases the sidewall area. Because the roughness of top/bottom surfaces is typically much less than that of the sidewalls, $Q$ may increase for smaller $H$. Interestingly, for very small $H$ values, the weaker modal confinement actually reduces the overlap between the optical mode and the top/bottom/sidewall surfaces. This effect is brought to bear on ultra-thin ($H<100$ nm), ultra-high-$Q$ ($>100$ million) SiN microresonators \cite{puckett2021422}. 

We work with a commercial foundry to design and fabricate microresonators on an $H = 160$ nm SiN platform. $Q$ measurements for TE$_0$ and TE$_1$ modes near $1060$ nm are shown in Fig. \ref{fig:one}(d). As expected, we observe higher $Q$ for increasing $RW$, with a maximum value of $Q=(18\pm2)\times 10^6$ for the TE$_0$ mode and $RW=3.5$ $\mu$m. In Supplementary Material, we show that our designs support TE$_0$ $Q>50$ million, using roughness parameters reported by commercial foundries.  

Kerr nonlinear photonics rely on wavelength conversion via four-wave mixing (FWM), in which one, two, or three input photons are converted to three, two, or one output photons at new frequencies \cite{boyd2020nonlinear}. Like all nonlinear optical conversions, phase- and frequency matching conditions (momentum and energy conservation) impose hard physical bounds on which frequencies can participate in FWM. In microring optical resonators, the phase matching requirement is

\begin{equation}\label{eq:pmatch}
        \sum_{n = 1}^{N}{m_n^{\rm{in}}}=\sum_{n = 1}^{4-N}m_n^{\rm{out}},
\end{equation}
where $N$ is a positive integer less than four and $m_n^{\rm{in (out)}}$ are integer azimuthal mode numbers of input (output) light. Likewise, the frequency matching requirement is

\begin{equation}\label{eq:fmatch}
    \sum_{n = 1}^{N}{\nu_n^{\rm{in}}}=\sum_{n = 1}^{4-N}\nu_n^{\rm{out}},
\end{equation}
where $\nu_n^{\rm{in (out)}}$ are input (output) frequencies of resonator modes, related to $m$ by $\nu(m) = \frac{m c}{n_{\rm{eff}}(m) L}$, where $c$ is the vacuum speed of light, $L$ is the resonator length, and $n_{\rm{eff}}$ is the effective refractive index of the mode. Notably, in devices where $n_{\rm{eff}}$ increases monotonically with $m$ (i.e., pure normal dispersion), Eqs. \ref{eq:pmatch} and \ref{eq:fmatch} cannot be satisfied simultaneously. This has separated high-$Q$ and Kerr-nonlinear regimes---the former utilizes small $H$ and large $RW$ to reduce scattering losses, but both of these design choices induce normal dispersion that is incompatible with the latter. 
\begin{figure*}[t!]
\centering
\includegraphics[width=\linewidth]{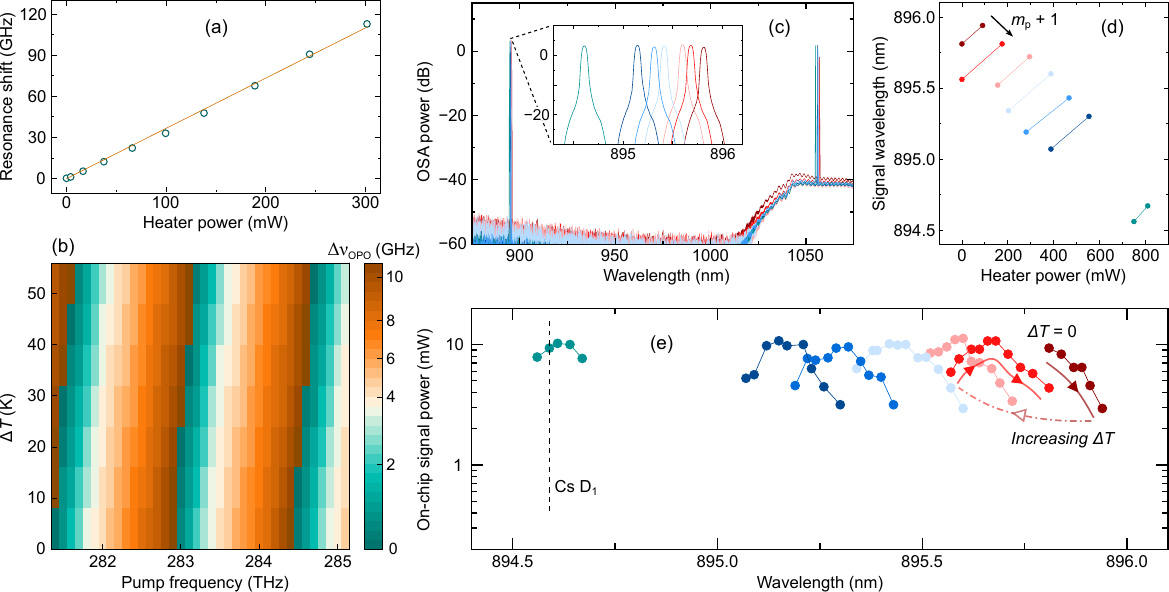}
\caption{\textbf{OPO tuning characteristics.} (a) Resonance frequency shift versus electrical power supplied to the on-chip heaters. (b) Simulated $\Delta\nu_{\rm{OPO}}$ versus pump frequency and temperature shift, $\Delta T$. OPO regions are shaded green. (c) OPO spectra obtained for different heater powers and pump frequencies. Different colors correspond to different pump azimuthal mode numbers, $m_{\rm{p}}$. Inset: zoomed into the signal band. $0$ dB is referenced to $1$ mW. (d) Range of OPO signal wavelength versus heater power. Moving down and rightward between lines corresponds to incrementing $m_{\rm{p}}$. Only wavelengths for which the on-chip signal power exceeds $2$ mW are included. (e) On-chip signal power versus signal wavelength. Each color corresponds to a different $m_{\rm{p}}$ value, as in (c) and (d). The arrowed path charts the OPO trajectory as heater power is increased: For a given $m_{\rm{p}}$ value, increasing heater power redshifts the signal wavelength.}\label{fig:three}
\end{figure*}

Figure \ref{fig:one}(e) illustrates the dispersion problem and its solution by modal phase matching, or hybrid-mode FWM. We plot simulated $n_{\rm{eff}}(m)$ spectra for the first four TE mode families in a device with $RW = 3$ $\mu$m. Each mode family exhibits strong normal dispersion. Another clear trend is that $n_{\rm{eff}}$ decreases with increasing transverse order, due to reduced mode overlap with the SiN waveguide. This feature adds a powerful design parameter. One can search through different mode families, selecting those with a combination of $\nu(m)$ that satisfy Eqs. \ref{eq:pmatch} and \ref{eq:fmatch}, to realize the desired FWM process, as illustrated by the FWM energy diagram inset to Fig. \ref{fig:one}(e). 

We instantiate this design principle using hybrid-mode OPO. In our case, OPO is an intraresonator FWM process that converts a monochromatic pump laser (injected into a single-mode bus waveguide that couples light to/from the microresonator) into a higher-frequency signal tone and a lower-frequency idler tone \cite{lu2025photonic}. In hybrid-mode OPO, the pump, signal, and idler modes may differ in their transverse order or polarization, as shown in Fig. \ref{fig:one}(a), while the bus waveguide supports only fundamental modes (i.e., all intraresonator lightwaves are outcoupled to a fundamental waveguide mode). Hybrid-mode OPO can promote greater output powers by suppressing competing nonlinear processes \cite{perez2023high, zhou2022hybrid}, but the potential to operate with foundry-friendly, high-$Q$ geometries has not been explored. 

We design a 1060-nm-pumped hybrid-mode OPO to target the Cs D$_{\rm{1}}$ transition near $894.6$ nm. In our design process, we parameterize the dispersion using the frequency mismatch spectrum, 
\begin{equation}\label{eq:mismatch}
    \Delta\nu = \nu_{\mu}^{fs} + \nu_{-\mu}^{fi} - 2\nu_{\rm{0}}^{fp},
\end{equation}
where $\nu_{\mu}$ is the frequency of a resonator mode with azimuthal mode number $m = \mu + m_{\rm{p}}$, where $m_{\rm{p}}$ is the azimuthal mode number of the pump mode. In Eq. \ref{eq:mismatch}, superscripts denote the spatial mode families (e.g., $fp = \rm{TE}_1$ denotes the pump mode), and Eq. \ref{eq:pmatch} is automatically satisfied. Hence, an OPO process conveyed by Eq. \ref{eq:mismatch} is efficient when $\Delta\nu \gtrsim 0$, where the approximation arises due to mode frequency shifts induced by the Kerr effect \cite{stone2022conversion}. We simulate $\nu(m)$ spectra for different mode families and $RW$, and find that the TE$_0$-TE$_1$-TE$_2$ mode configuration, where $fi-fp-fs$ denotes the idler, pump, and signal mode families, respectively, exhibits a $\Delta\nu$ spectrum well-suited to our target OPO. In Fig. \ref{fig:two}(a), we present $\Delta\nu$ spectra for various $RW$ values. Here, $m_{\rm{p}}$ is fixed, and the frequency axis corresponds to the signal mode frequency, $\nu_{\mu}$. The presence of $\Delta\nu$ zero crossings for all $RW$ illustrates the robustness of hybrid-mode OPO. The zero-crossing frequencies, written as $\nu_{\rm{s}}$ because they predict the OPO signal frequencies, are plotted versus $RW$ in the Fig. \ref{fig:two}(a) inset. Next, we choose $RW$ values that should produce a signal near the $894.6$ nm ($335$ THz) target wavelength (frequency). Based on Fig. \ref{fig:two}(a), we choose $RW$ values of $\{2.15$, $2.25$, and $2.35\}$ $\mu$m.     

\section{\label{sec:results}Results}
Photonic chips are fabricated by a commercial foundry and consist of a single, stoichiometric SiN device layer grown by low-pressure chemical vapor deposition on a $4$-$\mu$m-thick thermal oxide wafer, with a Ti+Al back-end-of-line process for on-chip heaters. With the expectation that OPO threshold and efficiency depend on resonator-waveguide coupling rates, we include a variety of coupling geometries in our designs \cite{stone2022efficient}. We record TE$_2$, TE$_1$, and TE$_0$ resonance lineshapes in the signal, pump, and idler bands, respectively. We search for near-critical coupling in the pump band (to minimize threshold power) and near-critical or slight overcoupling in the signal band to increase conversion efficiency while preserving a manageable threshold power. In Fig. \ref{fig:two}(b), we present transmission measurements for one such device, with $RW=2.35$ $\mu$m. The TE$_1$ pump mode is critically coupled with $Q\approx5.5$ million, and the TE$_2$ signal mode is overcoupled with $Q\approx4$ million and $\eta = 0.89$, where $\eta = \frac{Q_{\rm{tot}}}{Q_{\rm{c}}}$, $Q_{\rm{tot}}$ is the loaded quality factor, and $Q_{\rm{c}}$ is the coupling quality factor. The TE$_0$ idler mode is undercoupled with $Q\approx20$ million. From OPO theory, the expected pump-to-signal conversion efficiency, $CE = \frac{\eta}{4}\frac{\nu_{\rm{p}}}{\nu_{\rm{s}}}$, where $\nu_{\rm{p}}$ is the pump laser frequency, is approximately $20$ percent \cite{stone2022conversion, sayson2019octave}. 
\begin{figure*}[t!]
\centering
\includegraphics[width=\linewidth]{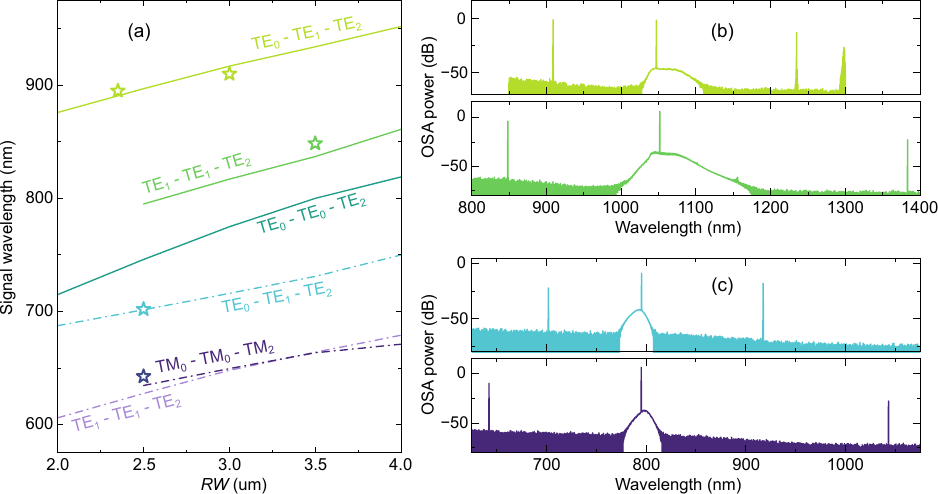}
\caption{\textbf{Hybrid OPO spectral coverage.} (a) Simulated signal wavelength versus $RW$ for different mode configurations (i.e., different sets of pump, signal, and idler mode families involved in the OPO process). Solid lines correspond to pumping near $1060$ nm, while dashed lines correspond to pumping near $795$ nm. Stars indicate experimentally measured values. (b-c) OPO spectra corresponding to the stars in (a), obtained through pumping near $1060$ nm and $795$ nm, respectively. $0$ dB is referenced to $1$ mW.}\label{fig:four}
\end{figure*}

We pump this device to generate OPO. The signal wavelength is $895.8$ nm, and the on-chip threshold pump power is approximately $10$ mW, though more pump power is required to maximize conversion efficiency \cite{stone2022conversion}. We generate $\approx10$ mW on-chip signal power using a $50$ mW pump. We present the output spectrum, recorded with an optical spectrum analyzer (OSA), in Fig. \ref{fig:two}(c). 

Next, we investigate OPO tuning via the on-chip heaters. We calibrate the heaters by the thermo-optic shift in resonance frequencies, from which we infer temperature changes, $\Delta T$. In Fig. \ref{fig:three}(a), we present measurements of this shift versus electrical heater power, observing the expected linear dependence. To understand how OPO depends on $\Delta T$, we calculate $\Delta \nu_{\rm{OPO}}$ as the smallest positive value in the $\Delta \nu (\mu)$ spectrum, plotting it versus $\Delta T$ and $\nu_{\rm{p}}$ in Fig. \ref{fig:three}(b). Two important relationships are apparent: First, $\Delta \nu_{\rm{OPO}}$ cycles as $\nu_{\rm{p}}$ is increased, forming narrow pump bands (shaded green) where OPO is possible. Second, increasing $\Delta T$ shifts these pump bands to higher frequencies, due to a temperature-dependent dispersion that arises primarily from the $n_{\rm{eff}}$ differences between pump, idler, and signal modes. This property of hybrid-mode OPO plays an important role in its frequency tuning. 

We experimentally verify this model and test OPO tuning by recording the OPO spectrum, shown in Fig. \ref{fig:three}(c), at various heater powers, re-adjusting $\nu_{\rm{p}}$ at each setpoint to follow the shifting pump resonance. We find that, to sustain OPO as $\Delta T$ is increased further, $m_{\rm{p}}$ must be incremented, as denoted in Fig. \ref{fig:three}(d), where each color designates a unique $m_{\rm{p}}$ value. This behavior is quantitatively consistent with our model. In Fig. \ref{fig:three}(e), we show how the OPO signal wavelength and power vary as $\Delta T$ is increased. At first, increasing $\Delta T$ redshifts the OPO signal, consistent with the redshift of resonator modes. However, when $m_{\rm{p}}$ is incremented to account for temperature-dependent dispersion, the OPO signal switches to a shorter wavelength. Increasing $\Delta T$ further repeats the cycle, shown by the arrowed path in Fig. \ref{fig:three}(e). Importantly, there is significant overlap in the range of signal wavelengths accessible at each $m_{\rm{p}}$ value, corresponding to gap-free wavelength tuning in this light source. Moreover, the tuning demonstrated here does not compromise efficiency --- the on-chip signal power stays above $5$ mW in most regions (an exception is near $\Delta T = 0$). For instance, we generate $>10$ mW on-chip signal power after adjusting the heater power and $m_{\rm{p}}$ to align the OPO signal wavelength with the Cs D$_1$ transition near $894.6$ nm, nearly $1.5$ nm away from its starting point.  

Finally, we explore wavelength accessibility in this hybrid-mode OPO platform. In Fig. \ref{fig:four}(a), we plot the simulated OPO signal wavelength versus $RW$ for different mode configurations, restricting ourselves to those with $>20$ percent FWM modal overlap \cite{lu2019efficient}. Three mode configurations and two standard pump wavelengths, $1060$ nm (solid lines) and $795$ nm (dashed lines), grant access to most signal wavelengths between $600$ nm and $950$ nm. This corresponds to a range of idler wavelengths between $850$ nm and $2$ $\mu$m. This platform is thus intriguing as a multi-wavelength source of laser light or entangled photon pairs. In Supplemental Material, we explore a larger set of $H$ and $RW$ values, showing that hybrid-mode OPO is compatible with many geometries. 

We experimentally verify our simulations by designing, fabricating (in collaboration with the commercial foundry), and testing microresonators with $RW$ values between $2.5$ $\mu$m and $3.5$ $\mu$m. We pump near either $1060$ nm or $795$ nm, generating several OPO states indicated with stars in Fig. \ref{fig:four}(a), and match them to our simulations. The corresponding optical spectra are presented in Figs. \ref{fig:four}(b-c). We observe remarkable agreement between the simulated and observed OPO signal wavelengths, including a $795$ nm-pumped OPO using transverse-magnetic (TM) mode families, which can exhibit higher $Q$ than TE modes when sidewall roughness is much greater than top-bottom surface roughness \cite{liu2022ultralow}. Importantly, the OPO efficiencies and output powers can be increased through careful design of the coupling geometry.  

\section{\label{sec:discussion}Conclusions}

We demonstrate that, compared to single-mode systems, hybrid-mode OPO designs are compatible with a much larger set of device geometries and foundry platforms. This allows OPO to be incorporated within multifunctional devices and, for example, bridges a longstanding gap between design approaches for Kerr nonlinear photonics - where ensuring suitable dispersion is paramount - and high-$Q$ microresonators that prioritize low optical losses. Hybrid-mode OPO is efficient and offers broad spectral coverage. Resonator dimensions accurately determine the signal wavelength, while temperature provides active, gap-free fine tuning. The design principle we use to implement modal phase matching for hybrid-mode OPO, namely that mode families with higher transverse orders possess larger eigenfrequencies, applies to any FWM process. By establishing this new framework, our work buttresses a new generation of high-$Q$ nonlinear devices manufacturable at scale. 

\section{Acknowledgements}
The authors thank Dr. Terence Blesin for helpful discussions. Effort sponsored by the U.S. Government through the Defense Innovation Unit (DIU) under Other Transaction number HQ0845-25-9-0037. The U.S. Government is authorized to reproduce and distribute reprints for Governmental purposes notwithstanding any copyright notation thereon. The views, opinions, and/or findings expressed are those of the author(s) and should not be interpreted as representing the official views or policies of the Department of Defense or the U.S. Government. J.S. and K.S. additionally acknowledge support from the NIST-on-a-chip program.

\section{Author contributions}
J.S. and K.S. developed the concepts and analyzed results. J.S. performed the measurements and wrote the manuscript, with input from all authors. S.F. performed and analyzed the simulations of microresonator quality factor and wrote the corresponding text. Devices were designed and fabricated with help from C.P., M.W., and G.K. 

\section{Competing interests}
J.S. and K.S. have non-provisional and provisional patents on Kerr OPO technologies. The authors declare no other competing interests.

\section{Data and materials availability}
All data supporting the main findings of this study are reported in the main text and supplementary materials. All other data related to this work are available upon reasonable request. 

\bibliography{bibliography}

\end{document}


\setlength{\textfloatsep}{12pt}

\title{Supplementary Material: Bridging high-\textit{Q} and Kerr-nonlinear photonics using modal phase matching}


\author{Jordan R. Stone}
\email{jordan.stone@nist.gov. This document is preliminary and is intended for peer review conducted by a journal or conference.}
\affiliation{National Institute for Standards and Technology, Gaithersburg, MD 20899}

\author{Saleha Fatema}
\affiliation{National Institute for Standards and Technology, Gaithersburg, MD 20899}
\affiliation{Whiting School of Engineering, Johns Hopkins University, Baltimore, MD 21218}

\author{Christopher V. Poulton}
\affiliation{Beacon Photonics, 3000 Wilson Blvd., Arlington, VA 22201}

\author{Michael G. Wood}
\affiliation{Beacon Photonics, 3000 Wilson Blvd., Arlington, VA 22201}

\author{Gordon A. Keeler}
\affiliation{Beacon Photonics, 3000 Wilson Blvd., Arlington, VA 22201}

\author{Kartik Srinivasan}
\affiliation{National Institute for Standards and Technology, Gaithersburg, MD 20899}
\affiliation{Joint Quantum Institute, NIST/University of Maryland, College Park, MD 20742}


\date{\today}

\begin{abstract}
In this document, we explain the scattering loss model used to understand quality factor measurements. We present simulation results supporting our central claim that hybrid-mode-OPO and ultra-high-$Q$ device geometries can coincide. We further detail the OPO threshold power and conversion efficiency calculations, as well as other materials and methods. Finally, we provide more simulation data around wavelength accessibility with hybrid-mode OPO.
\end{abstract}


\maketitle

\section{Scattering loss model for silicon-nitride microring resonators}

To quantify the effect of $Q$ with geometry, we calculate the scattering-limited intrinsic quality factor over a broad range of SiN waveguide widths and thicknesses using the volume-current scattering model developed by H\"{o}rmann et al \cite{hormann2023ab}. In this framework, a displacement of a dielectric boundary caused by surface roughness is represented by an equivalent polarization current. The radiation generated by these perturbation-induced currents is evaluated using the electromagnetic fields of the unperturbed guided mode. For each dielectric interface, the roughness is statistically characterized by its root-mean-square amplitude, $\sigma$, and its spatial autocorrelation length, $L_c$. 
The scattering calculation therefore accounts for two complementary effects: the electromagnetic field intensity and orientation at each dielectric boundary, which determine the sensitivity of a particular waveguide electromagnetic mode to boundary perturbations, and the roughness power spectral density, determined by $\sigma$ and $L_c$, which determines the spatial-frequency content available for scattering. While the latter effect depends on the available foundry and thus provides less control, the sensitivity of optical modes to the sidewall roughness can be optimized by engineering the geometry. 
The H\"{o}rmann framework is well suited to the present analysis because it explicitly incorporates the vector electromagnetic fields at each dielectric boundary and the statistical properties of the interface roughness. To verify our implementation, we first reproduced the results reported in Refs. \cite{horikawa2016low, fursenko2012characterization} using the corresponding waveguide geometries and roughness parameters. The calculated propagation losses/ values agreed with the published results within $<1~\%$ difference, providing an independent validation of our implementation before applying the model to the OPO device geometries considered here. 

While the original implementation by H\"{o}rmann et al. focuses on sidewall-roughness-induced scattering, we extend the model to independently calculate scattering contributions from the top and bottom interfaces. This  allows the contributions from the etched sidewalls and the top/bottom interfaces to be evaluated independently and therefore provides a direct connection between fabrication-induced side and top/bottom interface roughness, waveguide geometry, and resonator Q.
The calculated scattering power is converted to a propagation loss coefficient $\alpha_{\rm scatter}$. The corresponding scattering-limited intrinsic quality factor is $Q_{\rm scatter}=2\pi\times n_{\rm{g}}/(\lambda \times \alpha_{\rm scatter})$, where $n_{\rm{g}}$ is the group index of the mode, $\lambda$ is the wavelength, and $\alpha_{\rm scatter}$ is expressed as a power attenuation coefficient in units of 1/m.
Because scattering from physically distinct interfaces is treated as independent, the total scattering loss is obtained from $\alpha_{\rm scatter}=\alpha_{\rm scatter, top}+\alpha_{\rm scatter, bottom}+\alpha_{\rm scatter, sides}$
or equivalently, $1/Q_{\rm scatter}=1/Q_{\rm scatter, top}+1/Q_{\rm scatter, bottom}+1/Q_{\rm scatter, sides}$.
This decomposition is particularly useful for thin SiN waveguides because it reveals a transition from sidewall-limited to top/bottom-interface-limited operation as the waveguide becomes wider.

The roughness parameters of the fabricated interfaces were not independently measured for the devices considered here. We therefore use the experimentally measured intrinsic quality factors to obtain effective roughness parameters for the scattering model. We independently parameterize the etched sidewalls and horizontal SiN interfaces using RMS roughness amplitudes $\sigma_{\rm side}$ and $\sigma_{\rm tb}$, respectively. For simplicity, the upper and lower interfaces are assigned identical roughness statistics, $\sigma_{\rm top}=\sigma_{\rm bottom}=\sigma_{\rm tb}$ and $L_{c, \rm{top}}=L_{c, \rm{bottom}}=L_{c, \rm{tb}}$, respectively. This assumption reduces the number of free parameters but is not intended to imply that the physical top and bottom SiN interfaces possess identical roughness.
We calculated the resonator $Q_\text{scatter}$ over a range of roughness amplitude ($\sigma_{\rm side}$ and $\sigma_{\rm tb}$) and correlation lengths ($L_{c, \rm side}$ and $L_{c, \rm tb}$) and compared the results with the experimentally measured intrinsic Q. An effective parameter set of $\left[ \sigma_{\rm side}, L_{c, \rm{side}}, \sigma_{\rm tb}, L_{c, \rm{tb}} \right] = \left[4.9, 70, 0.44, 10\right]$ nm reproduces the measurements of the reference devices. To be in range of the expected roughness parameters for similar foundry-fabricated devices, $\left[\sigma_{\rm top}, \sigma_{\rm bottom}\right] = \left[0.2, 0.1\right]~\text{nm~to}~\left[0.45, 0.2 \right]$~nm \cite{puckett2021422}, we chose $\sigma_{\rm tb}=0.30$ nm while adjusting $\left[ \sigma_{\rm side}, L_{c, \rm{side}},  L_{c, \rm{tb}} \right] = \left[5.5, 60, 15\right]$ nm and subsequently examining the relative loss sensitivity of different waveguide geometries shown in Fig.~1(c) in main text.
These fitted values should be interpreted as effective scattering parameters rather than direct measurements of the physical interface roughness. The experimental intrinsic loss generally contains contributions from material absorption, radiation/bending loss, contamination, interface absorption, and other fabrication imperfections in addition to roughness-induced scattering. Because the calibration procedure attributes the measured loss entirely to scattering $(Q\approx Q_{\rm scatter})$, the extracted RMS roughness represents an effective upper-bound parameter within the present model. Consequently, the fitted values may exceed the actual physical roughness measured, for example, by atomic force microscopy (AFM) or scanning electron microscopy (SEM). 

\begin{figure*}[t!]
\centering
\includegraphics[width=\textwidth]{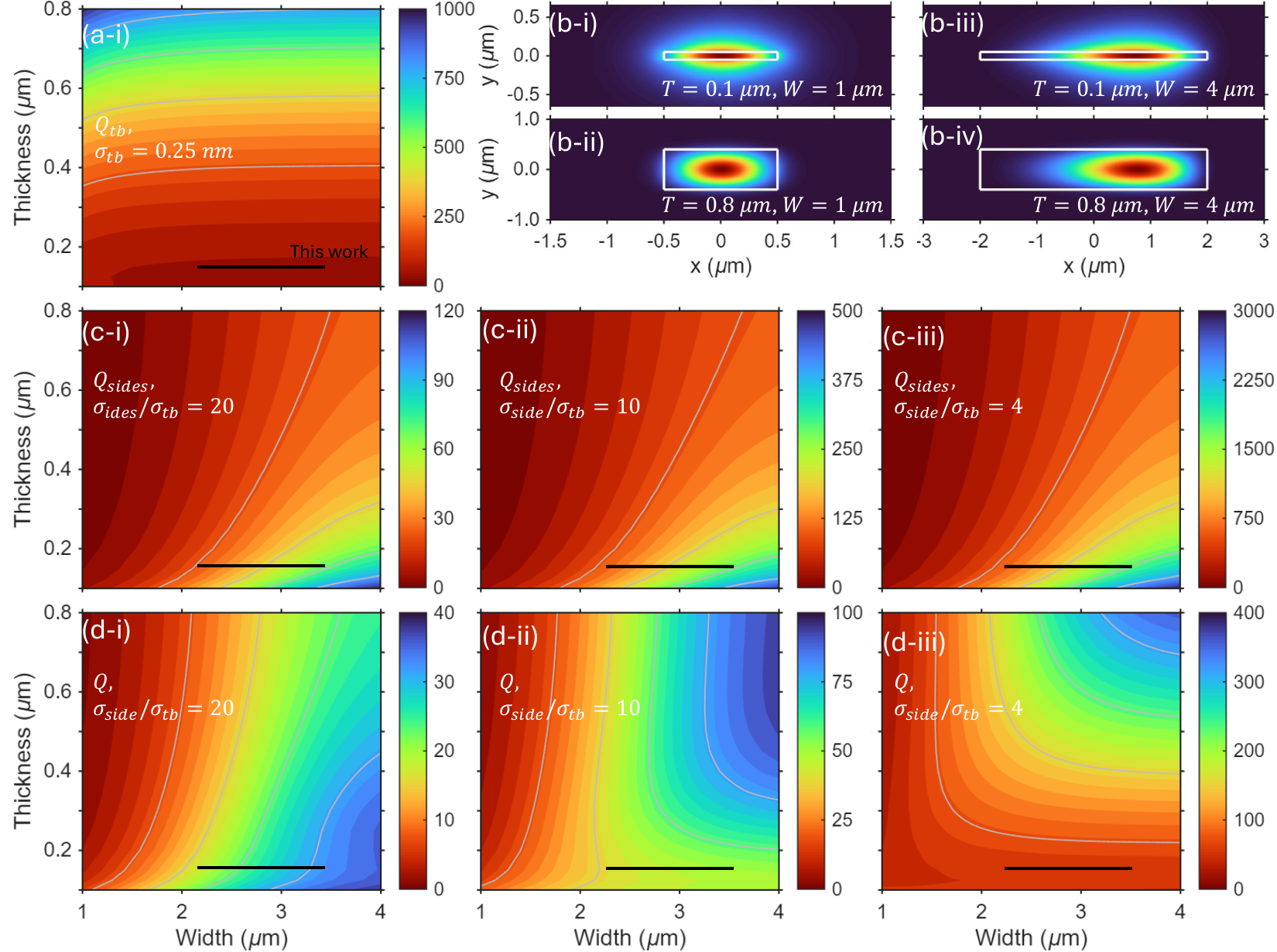}
\caption{\textbf{Geometry and roughness dependence of scattering-limited quality factor.} 
\textbf{(a-i)} Top/bottom-surface scattering-limited quality factor, $Q_{\mathrm{tb}}$ (million), as a function of waveguide width and thickness for $\sigma_{\mathrm{tb}} = 0.25$ nm.
\textbf{(b-i--b-iv)} Representative fundamental TE mode intensity profiles, $|E|^2$, for waveguide thicknesses $T = 0.1~\mu\mathrm{m}$ and $0.8~\mu\mathrm{m}$ and widths $W = 1~\mu\mathrm{m}$ and $4~\mu\mathrm{m}$, illustrating the evolution of modal confinement with waveguide geometry.
\textbf{(c-i--c-iii)} Sidewall-scattering-limited quality factor, $Q_{\mathrm{side}}$ (million), for $\sigma_{\mathrm{side}}/\sigma_{\mathrm{tb}} = 20$, 10, and 4, respectively.
\textbf{(d-i--d-iii)} Corresponding total scattering-limited intrinsic quality factor, $Q$ (million), including both sidewall and top/bottom-surface scattering.
The black horizontal lines indicate the waveguide-geometry range explored in this work.
Contours indicate constant-$Q$ values.
All calculations are performed at $\lambda = 1060$ nm with $\sigma_{\mathrm{tb}} = 0.25$ nm, $L_{c,\mathrm{side}} = 50$ nm, and $L_{c,\mathrm{tb}} = 10$ nm.
}
\label{fig:SM1}
\end{figure*}

Since the achievable intrinsic quality factor depends on both the optical mode distribution and the statistical properties of the interface roughness, we next investigate how the scattering-limited $Q$ evolves under different roughness conditions. Within the perturbative scattering model, we find that the calculated loss is substantially more sensitive to the RMS roughness amplitude, $\sigma$, than to variations in the correlation length, $L_c$, over the range relevant to our devices. We therefore fix $L_c$ and focus on the relative roughness of the etched sidewalls and the top/bottom SiN interfaces.
To illustrate the transition between different scattering regimes, we fix the top/bottom RMS roughness at $\sigma_{\mathrm{tb}}=0.25$ nm and consider sidewall roughness values of $\sigma_{\mathrm{side}}=\{5$, $2.5$, and $1$\} nm, corresponding to roughness ratios $\sigma_{\mathrm{side}}/\sigma_{\mathrm{tb}}=\{20$, $10$, and $4$\}, respectively. Figure~\ref{fig:SM1} shows the corresponding sidewall-limited, top/bottom-limited, and total scattering-limited $Q$ as functions of waveguide width and thickness. The total scattering-limited quality factor is calculated from the independent interface contributions according to $Q^{-1}=Q_{\mathrm{side}}^{-1}+Q_{\mathrm{tb}}^{-1}$.

The individual interface contributions exhibit distinct geometry dependences. Since $\sigma_{\mathrm{tb}}$ is fixed, $Q_{\mathrm{tb}}$ is identical for all three roughness cases and depends predominantly on waveguide thickness (Fig.~\ref{fig:SM1}(a)). As the thickness increases, the optical mode becomes less sensitive to perturbations of the horizontal interfaces (Fig.~\ref{fig:SM1}(b-i, b-ii)), resulting in a monotonic increase in $Q_{\mathrm{tb}}$ over the thickness range considered here. At very small SiN thicknesses ($<100$ nm), the reduction in scattering results from weak optical confinement, as a substantial fraction of the mode extends into the surrounding cladding and becomes less sensitive to perturbations of the SiN interfaces \cite{puckett2021422, liu2022ultralow}. We exclude this ultrathin regime from the present design comparison because the associated weak confinement also increases bend sensitivity and can require very large resonator radii to suppress radiation loss \cite{puckett2021422, liu2022ultralow}.
In contrast to $Q_{\mathrm{tb}}$, $Q_{\mathrm{side}}$ depends on both waveguide width and thickness. Increasing the waveguide width reduces the modal sensitivity to the etched sidewalls and therefore substantially increases $Q_{\mathrm{side}}$. 
Increasing the thickness has the opposite effect, increasing the interaction with the vertical interfaces and reducing $Q_{\mathrm{side}}$. Thus, thin and wide waveguides generally favor reduced sidewall scattering. 
In a microring resonator, however, the sidewall interaction is additionally influenced by the bending radius. At a fixed radius, increasing the waveguide width can enhance bend-induced modal asymmetry, displacing the optical mode toward the outer sidewall (Fig.~\ref{fig:SM1}(b-iii, b-iv)) and partially offsetting the reduction in sidewall scattering expected for wider waveguides \cite{roberts2022measurements}. For Fig.~\ref{fig:SM1}, a fixed radius of 250 $\mu$m us used; increasing the resonator radius reduces this bend-induced displacement \cite{roberts2022measurements}.
Although the magnitude of $Q_{\mathrm{side}}$ increases as $\sigma_{\mathrm{side}}$ is reduced, its overall dependence on waveguide geometry remains similar across the three cases considered here. The resulting total $Q$ is therefore governed by the relative strengths of the sidewall and top/bottom contributions.

For $\sigma_{\mathrm{side}}/\sigma_{\mathrm{tb}}=20$ (Fig.~\ref{fig:SM1}(c-i)), sidewall scattering dominates over most of the investigated design space. Consequently, $Q$ largely follows $Q_{\mathrm{side}}$, favoring thin and wide waveguides (Fig.~\ref{fig:SM1}(d-i)). 
At $\sigma_{\mathrm{side}}/\sigma_{\mathrm{tb}}=10$ (Fig.~\ref{fig:SM1}(c-ii)), the sidewall and top/bottom contributions become increasingly comparable. Increasing the waveguide width initially improves $Q$ by suppressing sidewall scattering; however, as the sidewall contribution decreases, scattering from the horizontal interfaces becomes increasingly important. The resulting $Q$ therefore exhibits a crossover between sidewall- and top/bottom-limited regimes (Fig.~\ref{fig:SM1}(d-ii)).
This crossover is even more pronounced for $\sigma_{\mathrm{side}}/\sigma_{\mathrm{tb}}=4$ (Fig.~\ref{fig:SM1}(c-iii)). With the improved sidewalls, $Q$ approaches the top/bottom-limited value as the waveguide is widened. The achievable $Q$ of thin, wide waveguides is therefore ultimately limited by the horizontal interfaces, and further improvement of the sidewalls provides insignificant returns (Fig.~\ref{fig:SM1}(d-iii)). In this regime, reducing the roughness of the top and bottom interfaces, for example through optimized interface preparation or planarization processes such as chemical-mechanical polishing (CMP) \cite{ji2017ultra}, provides a more effective route toward further improvement in $Q$.

Finally, the scattering-limited $Q$ is also wavelength dependent, as the modal confinement and field distribution at the waveguide interfaces vary with wavelength. While the calculations in Fig.~\ref{fig:SM1} are performed at a fixed wavelength of $1060$ nm, the absolute $Q$ values can change at other wavelengths. For example, at $1550$ nm wavelength, a $160$-nm-thick and $3.5~\mu\mathrm{m}$-wide SiN microring resonator yields $Q=\{60.74$, $123.21$, and $173.03$\} million for $\sigma_{\mathrm{side}}/\sigma_{\mathrm{tb}}=\{20$, $10$, and $4$\}, respectively.
\section{Wavelength accessibility with hybrid-mode OPO}
\begin{figure*}[t!]
\centering
\includegraphics[width=\linewidth]{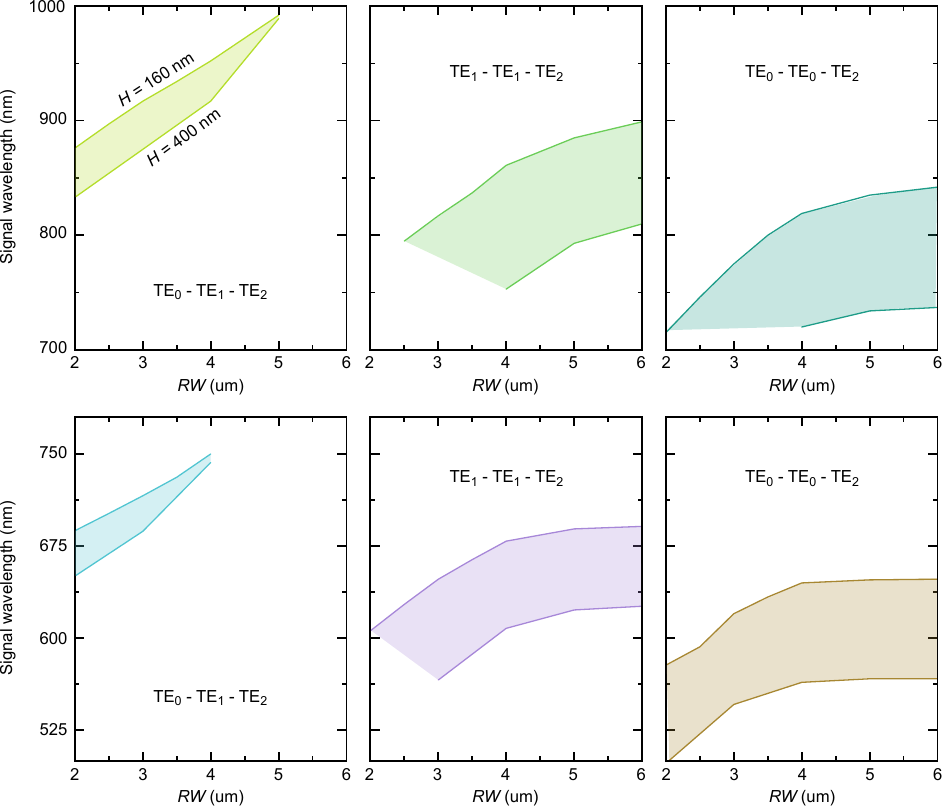}
\caption{\textbf{Wavelength accessibility with hybrid-mode OPO.} Simulated OPO signal wavelength versus $RW$ for $H = 160$ nm and $H = 400 $ nm and various mode configurations. The top row corresponds to $1060$-nm pumping, and the bottom row corresponds to $795$-nm pumping.}\label{fig:SM2}
\end{figure*}

Given these relationships between $Q$, $H$, and $RW$, it is important to assess the signal wavelengths produced by hybrid-mode OPO for different geometries, especially because $H$ will generally differ between SiN platforms. In Fig. \ref{fig:SM2} we plot the simulated signal wavelength versus $RW$ for three different mode configurations and two pump wavelengths ($1060$ nm and $795$ nm). Each graph shows two $H$ values, $160$ nm and $400$ nm, with the area between them shaded to portray the possible signal wavelengths for intermediate $H$. 

Interestingly, the increase in $H$ from $160$ nm to $400$ nm does not significantly shift the signal wavelengths. The curves maintain their shape while the signal shifts to shorter wavelength. Here, we only assess mode configurations involving TE$_0$, TE$_1$, and TE$_2$ families, since our experiments use relatively small $RW$ values. Larger $RW$ values could allow for the use of higher-order mode families, potentially expanding wavelength accessibility or providing another design parameter. Overall, modal phase matching offers a remarkably flexible design template for working across a wide area of geometric parameter space. In particular, for any $H$ value within the above range, most signal wavelengths between 720~nm and 990~nm (for 1060~nm pumping) and between 550~nm and 750~nm (for 795~nm pumping) are accessible.

\section{Materials and Methods}
\textbf{Dispersion simulations.} We use commercial finite-element-method software to calculate resonator eigenmodes, including their effective refractive indices, $n_{\rm{eff}}$ and wavenumbers, $m_{\rm{float}}$, at different frequencies. We then interpolate these frequencies onto the integer grid of azimuthal mode numbers, $m$, yielding the microresonator mode spectrum, $\nu(m)$, used in the main text. 

\textbf{Q measurements.} We record the microresonator transmission versus probe laser frequency, keeping the probe laser power below the onset of significant thermo-optic effects. The frequency axis is calibrated using a fiber Mach-Zehnder interferometer with $FSR\approx99$ MHz. Intrinsic and coupling linewidths are extracted by fitting the transmission lineshapes to a Lorentzian profile. 

\textbf{Calculations of OPO threshold and conversion efficiency.} First, we calculate the facet losses between the lensed fibers (used to inject and collect light to/from the bus waveguide) and the bus waveguide. Each facet uses the same inverse taper waveguide structure, so we assume identical losses at each facet. We perform these measurements in each relevant wavelength band (e.g., $1060$ nm and $895$ nm). We measure the optical power, $P_{\rm{in}}$ in the input lensed fiber, then subsequently measure the optical power, $P_{\rm{out}}$ in the output lensed fiber, calculating the facet loss as $\sqrt{P_{\rm{in}}/P_{\rm{out}}}$. 

To measure the OPO threshold power, we increase $P_{\rm{in}}$ to observe OPO. We then reduce $P_{\rm{in}}$ gradually, adjusting the pump frequency to stay on resonance, until OPO is no longer observed. We designate the value of $P_{\rm{in}}$ at which OPO disappears, after dividing by the facet loss, as the on-chip threshold power. Likewise, to calculate the conversion efficiency, we must estimate the on-chip OPO signal power. To do this, we inject light from a separate laser (with a wavelength in the signal band) into the bus waveguide, measuring $P_{\rm{out}}$. We then record the OSA power, where a network of optical components (necessary to our experiments) connects the output lensed fiber and OSA, and then calculate the optical loss between them. Hence, the on-chip signal power is the OSA power multipled by both the facet loss and this calibration factor. Then, the conversion efficiency is calculated as the on-chip signal power divided by the on-chip pump power.
\bibliography{Bibliography}